\documentclass[11pt,twoside]{article}
\usepackage{asp2010}

\resetcounters

\begin{document}

\title{Planets and X-rays: a radiation diet}
\author{J. Sanz-Forcada$^{1}$,
I. Ribas$^{2}$, 
G. Micela$^{3}$,
A. Pollock$^{4}$, 
D. Garc\'{i}a-\'Alvarez$^{5,6}$,
E. Solano$^{1,7}$, 
and 
C. Eiroa$^{8}$
\affil{$^1$Dpto. de Astrof\'{i}sica, Centro de Astrobiolog\'{i}a /
  CSIC-INTA, LAEFF Campus, P.O. Box 78, E-28691 Villanueva de la
  Ca\~nada, Madrid (Spain), jsanz@cab.inta-csic.es} 
\affil{$^2$Institut de Ci\`ences de l'Espai / CSIC-IEEC (Spain)}
\affil{$^3$INAF -- Osservatorio Astronomico di Palermo (Italy)}
\affil{$^4$XMM-Newton SOC, European Space Agency, ESAC (Spain)}
\affil{$^5$Instituto de Astrof\'{i}sica de Canarias (Spain)}
\affil{$^6$Grantecan CALP, La Palma (Spain)}
\affil{$^7$Spanish Virtual Observatory, Centro de Astrobiolog\'{i}a / CSIC-IEEC (Spain)}
\affil{$^8$Dpto. de F\'{i}sica Te\'orica, Universidad Aut\'onoma de Madrid (Spain)}
}

\begin{abstract}
According to theory, high energy emission from the coronae of cool
stars can severely erode the atmosphere of orbiting planets. To test
the long term effects of the erosion we study a large sample of
planet-hosting stars observed in X-rays. The results reveal that
massive planets ($M_{\rm p} \sin i > 1.5$~M$_{\rm J}$) may survive
only if exposed to low accumulated coronal radiation. The planet
HD~209458~b might have lost more than 1~M$_{\rm J}$ already, and other
cases, like $\tau$~Boo~b, could be losing mass at a rate of
3.4\,M$_{\oplus}$/Gyr. The strongest erosive 
effects would take place during the first stages of the stellar life,
when the faster rotation generates more energetic coronal
radiation. The planets with higher density seem to resist better the
radiation effects, as foreseen by models.  Current models need to be
improved to explain the observed distribution of planetary masses with
the coronal radiation received.
\end{abstract}

\section{Introduction}
The discovery of exoplanets has reached a point in which it is possible
to study more detailed properties of planets, such as mass
evolution. Once the planet is formed, and the original disc is
dissipated, the main agent interacting with the atmosphere should be
the high energy emission from the corona of the star, for late type
stars. The photons with $\lambda < 912$~\AA\ can ionize hydrogen atoms, 
assumed to be the main component of the atmosphere of giant
planets. The effects of X-rays ($\lambda\lambda$~1--100) and
Extreme Ultraviolet (EUV, $\lambda\lambda$~100--912) photons take
place at different heights in the atmosphere of the 
planet. While EUV photons mainly ionize the atoms in the upper
atmosphere, the X-rays penetrate deeper in the atmosphere. The
free electrons produce a cascade of collisions while the X-rays
photons are absorbed in the atmosphere \citep{cec09}. These
collisions heat the 
atmosphere yielding its ``inflation'' and
eventually the evaporation of a part of it. The gravity of the planet
act as a protection, trying to keep the atmosphere attached to the
planet. If we assume that the planet atmosphere is mainly composed by
hydrogen, and all the photons in the whole XUV range (X-rays+EUV) are
absorbed and contribute to the heating of the atmosphere, it is
possible to calculate the mass loss of the atmosphere by balancing the
losses with the planet gravity \citep{wat81,lam03}. An additional source of
mass losses takes place through the Roche Lobe for close-in planets
\citep{erk07} included in the formula through the variable $K$ ($K \le
1$). Some authors consider that evaporation actually takes
place at a point somewhere above the planet radius $R_{\rm p}$, at the
``expansion radius'' $R_1$ \citep{bar04}. The resulting formula
\citep{san10} is:

\begin{equation}\label{eq:general}
 \dot M=\frac{4 \pi \beta^3 R_{\rm p}^3 F_{\rm XUV}}{{\rm G} K M_{\rm  p}}
\end{equation}
where $\beta = R_1/R_{\rm p}$,
$F_{\rm XUV}$ is the X-ray and EUV flux at the planet orbit, 
and G is the gravitational constant. Although some heat may be absorbed
at $R_1$, most XUV flux should be absorbed below $R_{\rm
  p}$, where most of the atmosphere is enclosed. Thus we can safely
assume $\beta \simeq 1$. The formula can be simplified using the
mean density of the planet ($\rho$), and assume that $K \simeq 1$
(valid for most cases):

\begin{equation}\label{eq:massloss}
 \dot M=\frac{3 F_{\rm XUV}}{{\rm G} \rho}\, .
\end{equation}

Stars with spectral types ranging from mid-F to early-M have stellar
coronae with temperatures of $\sim$1~MK, exceeding 10~MK in the most
active cases. The high temperature material in the transition region
($\sim \log T=4-5.8$) and corona ($\sim \log T=5.8-7.4$) emits
copious X-rays and EUV flux. Fast rotators have a hotter corona,
resulting in a higher XUV flux. Since the younger stars have a faster
rotation, the XUV emission of the late type stars decreases with
time. The evolution of X-rays emission with age has been
studied for the Sun \citep[e.g.][]{mag87,ayr97,rib05} and extended to
G and M stars \citep{pen08a,pen08b}. A relation using late F to early
M stars has been calibrated by Garc\'es et al. \citetext{in preparation} 
in the X-rays band,
allowing us to calculate the age of the stars from its X-rays
emission, and to trace the time evolution of this emission. 
Current instruments allow only access to the X-rays band, but there is
little information on the EUV emission in stars other than the Sun,
limited to  
$\lambda\lambda$~100--400 for the best cases due to the absorption of
the radiation in the interstellar medium. 

The distribution of planet masses with current emission from the
coronae of the stars should reflect the accumulated effects
of the mass loss in the atmospheres of the planet over time. Besides, 
if we know the evolution of
the emission in the whole XUV band we 
should be able to trace the planet evolution, given an accurate
knowledge of the density of the planet, according to
Eq.~\ref{eq:massloss}. We carry out a program to observe the
X-rays emission of stars with exoplanets and evaluate these and other
effects. We have developed a database,
``X-exoplanets''\footnote{``X-exoplanets'' can be accessed from
  http://sdc.cab.inta-csic.es/xexoplanets}, 
that includes not only observations of the stars in X-rays, but also
synthetic spectra in the whole XUV (1--920~\AA) range \citep{san09},
available through the Spanish Virtual Observatory (SVO). We analyzed X-rays
data awarded to us either as P.I. or co-I., including an XMM-Newton Large
Observation  
(proposals \#02065304, \#02000001, \#05510229), and archival XMM-Newton
and Chandra data. The scientific results \citep[Sanz-Forcada et
al., in preparation]{san10} are outlined in next section.

\begin{figure}[t]
\center
\plotone[width=\textwidth,clip]{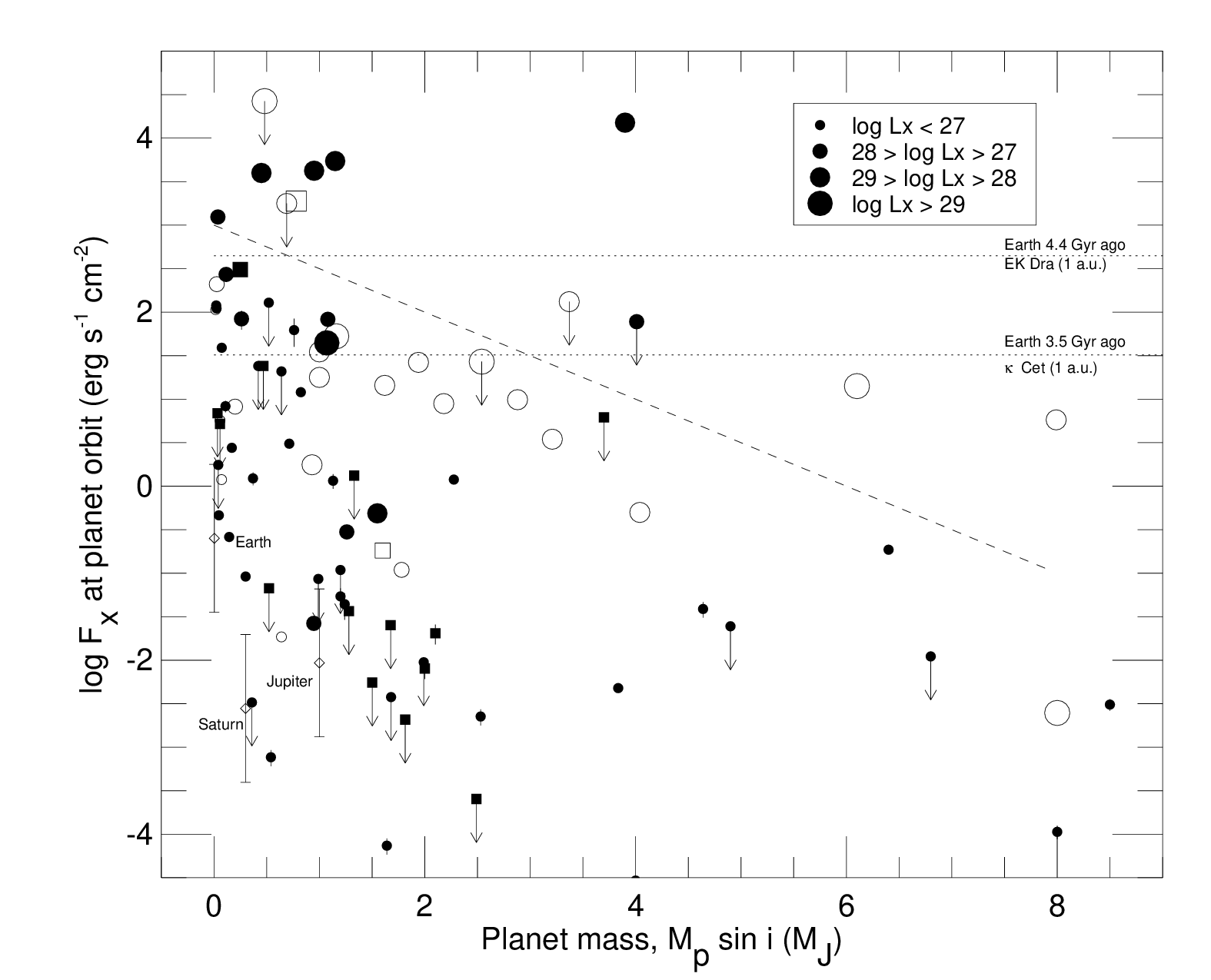} 
\caption{\label{fig1}Distribution of planetary masses ($M_{\rm p} \sin
  i$) with X-ray flux at the planet orbit. Filled symbols (squares for
  subgiants, circles for dwarfs) are XMM-Newton and Chandra
  data. Arrows indicate upper limits. Open symbols are ROSAT data
  without error bars. Diamonds represent Jupiter, Saturn, and the
  Earth. The dashed line marks the ``erosion line''. Dotted lines
  indicate the X-ray flux of the younger Sun at 1 a.u. Adapted from
  \citet{san10}.
}
\end{figure}

\begin{figure}[t]
\center
\plotone[width=\textwidth]{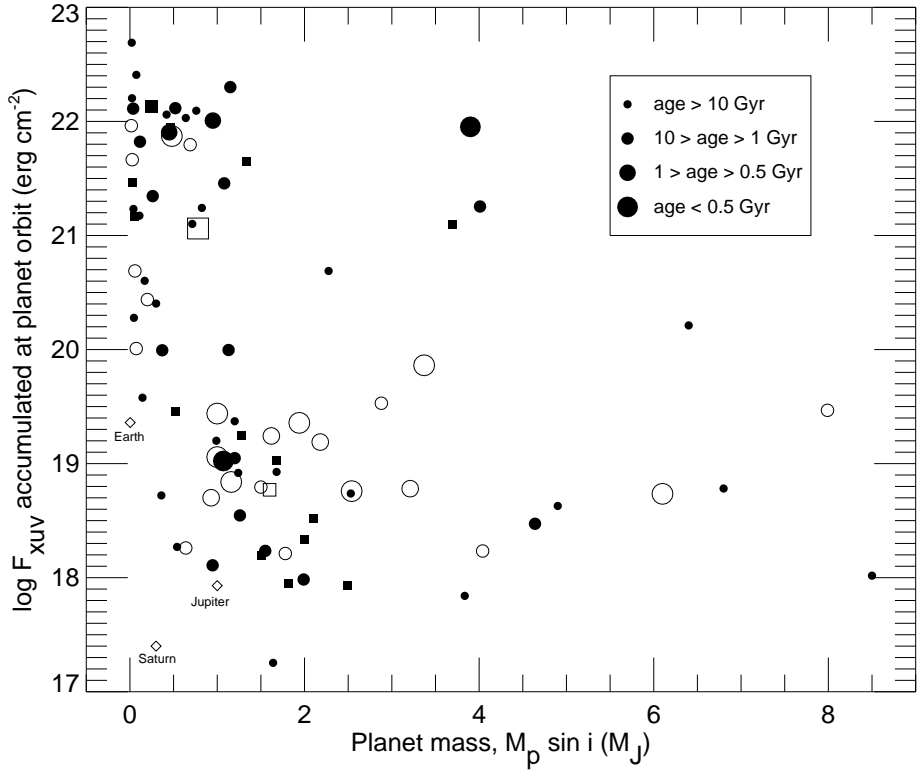} 
\caption{\label{fig2} Distribution of planetary masses ($M_{\rm p}
  \sin i$) with the X-ray flux accumulated at the planet orbit since
  an age of 20 Myr to the  present (see text). Symbols as in
  Fig.~\ref{fig1}. 
}
\end{figure}

\section{Results}
Direct XUV observational data of stars with exoplanets can be acquired
only in the X-rays band with present instruments. Since X-rays and EUV
are formed in the same region and temperatures, the trends observed in
X-rays should be extrapolative to the whole XUV range. We have
calculated the X-rays flux by fitting the spectra of all the stars
with exoplanets observed with XMM-Newton or Chandra, and complemented
these data with ROSAT observations with S/N$>$3. A total of 101
planets in 88 stars are included in the sample. We can interpret the
results keeping in mind Eq.~\ref{eq:massloss}. The distribution of
$F_{\rm XUV}$ with masses (Fig.~\ref{fig1}) reveal a clear separation
that seems to be related to mass\footnote{Planets orbiting giants
  and cataclysmic variables were excluded from the analysis.}. We
plotted a line (``erosion line'')
that roughly follows this separation. This line is not based on any
previous assumption or physical law. The Solar System planets are
included for comparison, with vertical segments indicating the
variations over the solar cycle. We also include the radiation level
arriving at the Earth when the life appeared, 3.5 Gya, and
when the Sun had an age of only 1 Gyr. We used the solar young analogs
\citep{rib05} 
$\kappa^1$~Cet and EK Dra to mimic the flux at the Earth's orbit in
the past. Fig.~\ref{fig1} can be interpreted in the same manner as the HR
diagram: the absence of planets in a given area of the diagram implies
that they spend little time in that stage. In this sense the
distribution seems to indicate that the planets suffer heavy erosion
in the first stages until they are below the ``erosion line'', when
either the stellar XUV flux decreases and/or the gravity of the planet
protects the planet from mass loss, slowing down the erosion. Planets
positioned initially at further distances will not suffer substantial
mass loss. We can also divide the diagram in four quadrants based on
the mass (at
1.5~M$_{\rm J}$) and the X-ray flux at the planet orbit (at $\log
F_{\rm X}$=2.15). Only 1 out of 12 of the planets receiving high flux
have high mass, while 46 out of 89 (a 52\%) of the 
planets receiving lower fluxes have a high mass. Therefore it is clear
the absence of high mass planets suffering high flux levels.   

\begin{figure}[t]
\plotone[angle=90,width=0.95\textwidth]{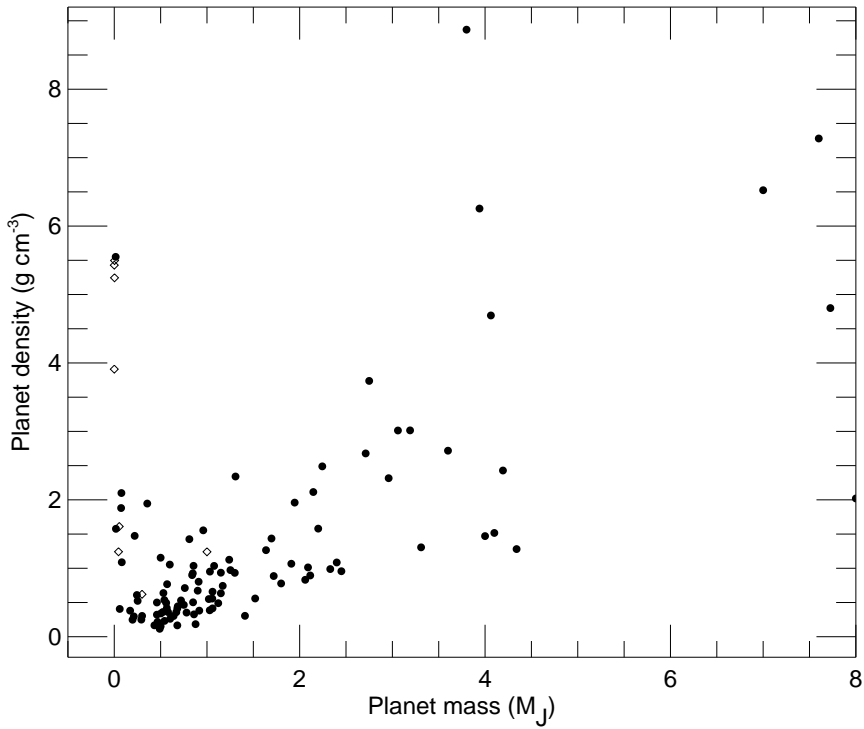}
\caption{\label{fig3} Density of the 103 planets of known
  radius (17 Nov 
  2010) with $M_{\rm p} < 8$\,M$_{\rm J}$ (filled circles). Diamonds
  represent the Solar System planets. 
}
\end{figure}

To test the effects over time we have also used the dependence of
X-rays emission with time. We need also the EUV evolution with time,
but the only estimates made to date are valid just for the Sun
\citep{rib05}, and a relation applicable to all late type stars is
necessary. The first step needed is to calculate the EUV flux. The
process is rather complex and it is better explained in
\citet{san09} and Sanz-Forcada et al. \citetext{in preparation}. 
We use the spectral information in X-rays to
derive a coronal model, extrapolate the model to the transition
region if no UV lines are available\footnote{The model of the
  temperature structure in the transition region is typically calculated from
  UV lines, formed at lower temperatures than those in X-rays.}, and
predict the spectral energy distribution in the whole EUV range using
the atomic model APED \citep{aped}. We have derived a relation between X-rays
and EUV luminosity ($L_{\rm EUV} \sim L_{\rm X}^{2.65}$) that we use
to calculate the EUV flux of the stars with ROSAT data. 
We also calculated the age of the stars using
their X-rays luminosity, and constructed a relation in the EUV range
with the best data of our sample ($\log L_{\rm EUV} \simeq 29.22 -1.41\,
\log \tau$, where $\tau$ is the age in Gyr).  Fig.~\ref{fig2} shows
the accumulated flux at the orbit of the planet, considering the time
between an age of 20~Myr and the present. The observed distribution
with mass would be a direct indication of the mass lost by the planets
since an age of 20~Myr, assuming that all mass losses are included in
Eq.~\ref{eq:massloss}, and the density is the same for all
planets. The distribution shows that only three planets with
$M_p>1.5M_{\rm J}$ have survived an XUV flux of more than
$10^{21}$\,erg\,cm$^{-2}$: for two of them there are alternative
explanations \citep{san10} and the third, $\tau$~Boo~b, is a young
object that might still be suffering heavy erosion. Alternatively, if
any of these planets has a rocky composition ($\rho
\ga$5\,g\,cm$^{-3}$) the erosion effects should be much lower, if any.

\begin{figure}[t]
\plotone{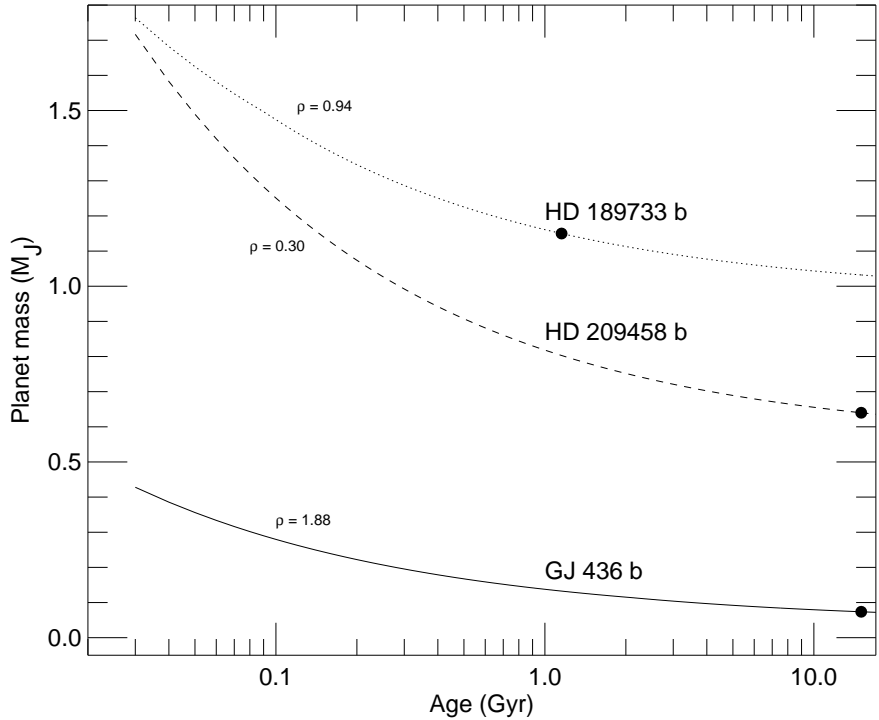} 
\caption{\label{fig4} Planetary mass
  evolution as an effect of the XUV radiation. Mass losses include
  evaporation by coronal radiation and losses through the Roche
  Lobe. Planet mean density (in g\,cm$^{-3}$) is indicated, as well as
  the current 
  stellar age of the planet,
  calculated using the X-ray luminosity (see text). 
}
\end{figure}

The second important variable in  Eq.~\ref{eq:massloss} is the planet
mean density. A more dense planet is better protected against
evaporation. Ideally our analysis should concentrate on stars with
known $L_{\rm X}$ and planets with known density. Only four planets
have both magnitudes known, and one of them, 2M1207~b is too far from
the star to suffer mass losses. Instead we can test the whole
population of planets with known density, a total of 103 planets. Most
of these planets have been detected trough the transits technique,
therefore having a strong bias towards close-in planets, those more
exposed to high XUV flux. We do not find massive planets with small
density among this population (Fig.~\ref{fig4}). 
This might indicate that low density
planets with high mass have suffered a quick erosion taking them to
the population with jovian masses, or increasing their density because
of the lower atmosphere-to-nucleus ratio. 

Finally we have made a calculation of the mass evolution for the
mentioned three planets with known density and $L_{\rm X}$, including
the calculated $L_{\rm EUV}$, according to
Eq.~\ref{eq:massloss}, and including in this case the mass losses
through the Roche Lobe expressed in Eq.~\ref{eq:general}. 
Fig.~\ref{fig4} shows this evolution assuming
that density has been roughly constant over time and no other effects
toke place. The case of HD~209458~b is remarkable, with more than
1~M$_{\rm J}$ lost to date. 
This work has considered only a rather simple model with the thermal
losses in the atmosphere, but non-thermal 
losses should be included in the future, together with other
smaller effects that may be necessary to explain the distribution
observed in Fig.~\ref{fig1}, as described in \citet{san10}.

\acknowledgements JSF and DGA acknowledge support from the
Spanish MICINN through grant AYA2008-02038 
and the Ram\'on y Cajal Program ref. RYC-2005-000549. 
IR acknowledges support from the Spanish MICINN via grant
AYA2006-15623-C02-01.
This research has made use of the NASA's High Energy
Astrophysics Science Archive Research Center (HEASARC) 
and the public archives of XMM-Newton and Chandra.


\end{document}